\documentclass[twoside,a4paper,12pt]{article}
\usepackage{graphicx}
\usepackage{mathtools}
\usepackage[pdfencoding=auto, psdextra]{hyperref}
\hypersetup{
    colorlinks=true,
    linkcolor=blue,
    filecolor=magenta,      
    urlcolor=cyan,
    pdftitle={A Practical Overview of Quantum Computing:  Is Exascale Possible?},
    pdfpagemode=FullScreen,
}
\usepackage[english]{babel}
\usepackage{gensymb}
\usepackage{bookmark}
\usepackage{PRIMEarxiv}
\usepackage[utf8]{inputenc} 
\usepackage[T1]{fontenc}    
\usepackage{booktabs}       
\usepackage{amsfonts}       
\usepackage{nicefrac}       
\usepackage{microtype}      
\usepackage{fancyhdr}       

\def\bottom{\perp}
\def\Norm{\mathop{\rm Norm}\nolimits}
\title{A Practical Overview of Quantum Computing: is Exascale Possible?
}

\author{James H. Davenport\thanks{Author list in alphabetical order.  See \url{https://www.ams.org/profession/leaders/CultureStatement04.pdf}}\\
\textit{Department of Computer Science}\\\textit{University of Bath}\\ Bath, UK\\
\texttt{masjhd@bath.ac.uk}
\And
Jessica R. Jones\footnotemark[\value{footnote}]\\
\textit{HPC/AI EMEA Research Lab}\\\textit{Hewlett Packard Labs}\\Bristol, UK\\
\texttt{j.r.jones@hpe.com}
\And
Matthew Thomason\footnotemark[\value{footnote}]\\
\textit{Hewlett Packard Enterprise}\\Bristol, UK\\
\texttt{matthew.thomason@hpe.com}
}
\date{May 2023}

\begin{document}

\maketitle

\begin{abstract}
Despite numerous advances in the field and a seemingly ever-increasing amount
of investment, we are still some years away from seeing a production quantum
computer in action.  However, it is possible to make some educated
guesses about the operational difficulties and challenges that may be encountered in
practice.  We can be reasonably confident that the early machines will be hybrid, with
the quantum devices used in an apparently similar way to current accelerators such as FPGAs
or GPUs.  Compilers, libraries and the other tools relied upon currently for
development of software will have to evolve/be reinvented to support the new technology, and
training courses will have to be rethought completely rather than ``just'' updated alongside them.

The workloads we are likely to see making best use of these hybrid machines
will initially be few, before rapidly increasing in diversity as we saw with
the uptake of GPUs and other new technologies in the past.  This will again be helped by the increase in the number of supporting libraries and development tools, and by the gradual re-development of existing software, to make use of the new quantum devices.

Unfortunately, at present the problem of error correction is still largely unsolved, although there have been many advances.
Quantum computation is very sensitive to noise, leading to frequent errors during
execution.
%
Quantum calculations, although asymptotically faster than their equivalents in ``traditional'' HPC, still
take time, and while the profiling tools and programming approaches will have to change drastically, many of the skills honed in the current HPC industry will not suddenly
become obsolete, but continue to be useful in the quantum era.
\end{abstract}

\keywords{HPC \and quantum computing \and exascale \and hybrid computing \and distributed computing}

\section{Introduction}
\setcounter{footnote}{0} 
At the present time, no one really knows what a production quantum computer
might look like, with many designs still being actively developed and researched.  It seems clear
however, based on the state of current HPC, that there will be a concerted
push from both industry and research towards larger and larger machines.  Taking
the current exaflop HPC systems as an example, it does not seem unreasonable
to expect an exa-op quantum computer to come into existence before the end of
the current century.  We may even see a usable hybrid machine, widely accepted to be the most likely early large-scale machine built with quantum devices \cite{ruefenacht2022bringing}, by 2050, but the
scale is likely to be far lower than we might want for most general-purpose quantum
computations.
\subsection{Parallelism}
It is important to note that parallelism is not as easy to
achieve in quantum algorithms as it is in more traditional HPC, the latter
having the benefit of many more years of active research.

An easy way to see this is via a simple look at AES-128 (reality is more complicated: see \cite{davenport2021improvements} for a recent analysis). Sequential search to find the key, given a matching plaintext/ciphertext pair, takes $2^{128}$ ``is this the right key'' operations, which we assume is done by some ``oracle'' code. In the classical world, this can trivially be divided across, say, $2^{64}$ computers each testing $2^{64}$ possibilities, even though this is wildly unfeasible. In the quantum world, Grover's algorithm \cite{grover1996fast} would require roughly $2^{64}\approx 16\cdot 10^{18}$ executions of the ``is this the right key'' oracle. At 1ns/oracle (pretty optimistic!) this is 500 years. The optimist says ``split this across a thousand computers, and it takes half a year''. But this means each computer is analysing $2^{128}/1000\approx 2^{118}$ possible keys. Grover's algorithm then requires $2^{59}$ oracle executions, which is 16 years, not the half-year the optimist expected. The problem is that the optimist was trying to take 1000-way parallelism \emph{inside} the $\sqrt{\strut}$ of Grover, and this isn't how quantum/parallelism interact. To get down to half-year, we would need to split the computation across a million computers, not a thousand.

Parallelism in quantum computation has been hypothesised in \cite{fowler2013timeoptimal} \cite{fowler2012surface} \cite{gidney2019flexible} over \emph{large} quantum computers, but not the hybrid systems of smaller networked quantum devices we anticipate being available in the nearer term.

A machine may be an exaflop machine, doing $10^{18}$ operations/second, but that doesn't mean it can perform a single operation in an attosecond. Rather, it is (roughly speaking) doing $10^9$ operations in parallel, each taking a nanosecond. It is currently not clear how that $10^9$-parallelism might translate to a quantum model of computation.
\subsection{Error Correction}\label{sec:qec}
In addition to the difficulties in exploiting parallelism, qubit fidelity is
not yet as high as it needs to be for most areas of scientific research to
exploit this new technology.  What's more, it probably never will be unless the problem of quantum error correction (QEC) is solved.

The current generation of quantum devices are
described as noisy, intermedi\-ate-scale quantum (NISQ) devices because of the
many hardware challenges they present, both in their size (low numbers of
qubits) and high error rates.  Though a classical digital computer is ultimately reliant on analogue voltages/charges, these are always being regarded as 0/1, i.e. effectively being reshaped. This can't be done with a quantum computer, as we need to preserve the quantum nature of the signal. See \cite{Amicoetal2019a} for the difficulties with Shor's algorithm (Section \ref{sec:Shor}).

Even the smallest of classical circuits, including tiny novel versions\footnote{It is worth noting that, due to their size, very small classical circuits must mitigate against quantum effects, thus limiting their usefulness.} 
such as
\cite{kim2012one}, may have hundreds of atoms in a single cell, whereas by
definition a quantum computer is reliant on a single photon/electron/\dots, and its basis state which could be, for example, its atomic spin state (for an atomic particle, such as an electron), photonic
polarization (in the case of a photon), charge state of a superconducting system, or the electronic state of an ion,
depending on the underlying technology employed. This means that the chance of encountering
an error is much higher, as is the susceptibility to external
noise.  The greater the ``depth'' (that is, the longer the sequence of gates in the quantum circuit), the greater the error propagation will be, so thus far only shallow quantum circuits have produced meaningful results.  This (along with the limited availability of higher qubit count hardware) has limited scientific use to very small problems, such as \cite{peruzzo2014variational}.

A classical circuit cannot directly correct a quantum error, so we
would appear to be reliant on error-prone quantum circuits to correct
error-prone circuits, which looks like a conundrum.

Substantial research is going on in this area, of course, resulting in
several different approaches for error mitigation, as well as bringing us ever closer to the holy grail of full quantum error correction.  Many of these mitigation techniques are directed at
specific technologies, such as zero-noise extrapolation \cite{giurgica2020digital},
probabilistic error cancellation (PEC) \cite{van2023probabilistic}, and
noise-tailoring techniques such as randomised compilation
\cite{hashim2021randomized}, all of which attempt to mitigate gate errors.  It is worth considering that quantum error correction, a good overview of which can be found in \cite{devitt2013quantum}, as well as most mitigation techniques, dramatically increase the number of qubits required for a single calculation\cite{fowler2012surface} \cite{campbell2019}, thus making it even more difficult to employ with current quantum hardware.

Nevertheless, it will require \emph{major} improvements to
get significant improvements here, and it is likely that quantum computers,
even with error-correction, will be less reliable than classical computers for
a while.  This means that algorithms will have to be more fault tolerant on
these early quantum computers than has been required on classical computers
for some decades. 

With such a thought in mind, let us consider an example use case for this type
of machine.  We begin with one of the most famous and popular examples.  For a comprehensive list of quantum algorithms, see \cite{quantumAlgorithmZoo}.

%
\section{Example usage: Factoring an \textit{N}-bit number}

Let's assume we are trying to factor an $N$-bit number, as required for the RSA cryptosystem. If the reader objects that RSA has been replaced by elliptic curve cryptography, we should note that Shor's Algorithm can also find elliptic curve discrete logarithms, and therefore breaks ECDSA and elliptic curve Diffie--Hellman as well.  Since the classical best algorithm for elliptic curve discrete logarithms is also based on the number field sieve, it is likely that the approach in \S\ref{sec:Bernstein} will generalise here.

\subsection{Classical Approaches}\label{sec:NFS}
The current approach for factoring $N$-bit numbers is the ``General Number Field Sieve'', or GNFS --- see \cite{buchmann1994implementation}.  A very crude sketch of the algorithm is as follows.
\begin{enumerate}
    \item[Setup]Choose a degree $d$, a number $m\approx N^{1/d}$ and an irreducible polynomial $f$ of degree $d$ such that $f(m)\equiv 0 \pmod N$. Typically we choose $d$, then $m\approx N^{1/d}$, write $N$ in base $m$, regard the ``digits'' as the coefficients of $f$, and $f$ is almost certainly irreducible. There is more in \cite{Murphy1999}.  Let $\alpha$ be a root of $f$ (regarded as an algebraic object: we don't care about a numeric approximation to it).
    \item[Sizing]Choose a bound $B$, and state that an integer whose prime factors are at most $B$ is \emph{$B$-smooth}. For an algebraic number $\beta$, we will say that it is \emph{$B$-smooth} if its norm $\Norm(\beta)$ is $B$-smooth. Let $p_1,p_2,\ldots,p_{\pi(B)}$ be the primes $\le B$. $\pi(B)$, the number of primes $\le B$, is roughly $\frac B{\log B}$.
    \item Search a large region of $(a,b)$ pairs with $-M\le a\le M$, $0\le b\le M$ for pairs $(a,b)$ such that both $a+mb$ and $a+\alpha b$ are $B$-smooth. So $a+mb=\prod p_i^{k_i}$ and $\Norm(a+\alpha b)=\prod p_i^{l_i}$. We need (slightly more than) $2\pi_B$ such pairs, and this requirement determines $M$.
    \item Choose (essentially linear algebra modulo 2) a subset $(a_j,b_j)$ of the pairs found in the previous step such that, when we write $\prod_j\Norm(a_j+mb_j)=\prod p_i^{k_i}$ and $\prod_j(a_j+\alpha b_j)=\prod p_i^{l_i}$, both the $k_i$ and the $l_i$ are all even. Note that the linear system in the $k_i$ and $l_i$ is very sparse: most primes won't occur in any particular factorization of $a_j+mb_j$ or $a_j+\alpha b_j$. Hence we do a certain amount of pre-processing (as in \cite{HoltDavenport2003}) to reduce the size $M_1$ of the matrix from the original $2\pi_B$, and then usually use Coppersmith's block Wiedemann \cite{Coppersmith1994b,Thome2002}, whose running time is $O(M_1^2)$ to solve the system.
    \item[Result]This, roughly speaking lets us find $x,y$ such that $x^2\equiv y^2\pmod N$. Then $\gcd(x-y,N)$ is a factor, with high probability non-trivial, of $N$.  If by any chance the factors are trivial, we find a different subset.
\end{enumerate}
Steps 1 and 2 consume the vast majority of the running time (though \cite{Murphy1999} suggests spending 3\% of the total time budget on the setup step). There is a trade-off, controlled by $B$: making $B$ larger makes it more likely that a pair $(a,b)$ will satisfy the smoothness criteria, and hence the searching in step 1 can be done over a smaller region (this outweighs the fact that $\pi(B)$ increases). However, the size of the linear system to be solved in step 2 increases.

For the best choice of $B$, the theoretical time is conjectured (i.e. proven subject to standard number-theoretic conjectures) to be 
\begin{equation}\label{eq:GNFS}
\exp\left(\left((64/9)^{1/3}+o(1)\right)(\log N)^{1/3}(\log\log N)^{2/3}\right).
\end{equation}
Note that $(64/9)^{1/3}\approx 1.923$.
\subsection{Shor's Algorithm}\label{sec:Shor}
The generally
accepted method of doing this on a quantum computer is Shor's algorithm
\cite{Shor1994a}, which will reduce the problem from one requiring exponential
time to one that can be solved in polynomial time.  At the time of its
publication, and often repeated, it was explained that the first true quantum
computers would render current encryption algorithms obsolete. Having said that, Shor's algorithm does not seem easy in today's world. The largest number factored \emph{with Shor's algorithm} on a quantum computer seems to be 21 \cite{MartinLopezetal2012a} in 2012. 
In 2019, an attempt \cite{Amicoetal2019a}
to factor 35 failed due to the accumulated errors. Hence statements like ``in practice Shor's algorithm \dots'' have to be taken with a peck of salt. This point is made forcibly in \cite{Smolinetal2012a}, who point out that the computation in \cite{MartinLopezetal2012a} (and the previous factorisation of 15) were ``compiled'' (so far so good), \textit{but} the compiler ``knew the answer'', which is not acceptable. \cite{Smolinetal2012a} goes on to argue that, because of this, just saying ``how large a number did you factor'' is not a measure of quantum computing progress.

The first problem that must be resolved is the number of qubits
required, which increases linearly with the number of bits in $N$. This means that, realistically,
a very large quantum computer will be needed to break current RSA-based
encryption.  For example, if a user of our quantum supercomputer were to be
attempting to break RSA-2048, using Shor's algorithm, they would require
around 4100 error-free qubits.

The second problem is a practical one of fault tolerance.  Shor's algorithm
requires a \emph{perfect} run or \emph{flawless error correction} to be successful.
In practice, Shor's algorithm requires $2N+O(1)$ \emph{perfect}
quantum bits, and $O(N^3)$ ``depth''\footnote{Variants using faster multiplication algorithms than ``schoolboy'' can reduce the depth at the cost of increasing the number of qubits required.}.  It also must run for around $10^{10}$ cycles without errors to factor RSA-2048.  Yes, we can run the same calculation repeatedly, but at some
point at least one \emph{flawless} run is required.  It is not possible to
combine two imperfect runs.

Such a
requirement should immediately cause concern in anyone already familiar with
the day to day difficulties of running a supercomputer centre, especially when
combined with the large number of required qubits.  ``Traditional'' computers
have good, rapid error detection and correction in both hardware and software.  They also
have known component failure rates, and, while hardware failure prediction
continues to be an area of active research, careful planning of hardware
maintenance alleviates most difficulties while minimising downtime.

In contrast, quantum hardware failure rates are rarely shared by manufacturers, and without a production-ready quantum machine running for several months, no one knows what the hardware failure rates, and consequently the maintenance requirements, may be.  Note that there is no known way to do checkpoint/restart for a genuinely quantum computation.  Thus any sudden downtime is likely to result in the loss of all running work up to that point.


A particular mapping of Shor's algorithm onto a hypothetical physical realisation can be found in \cite{GidneyEkera2021a}: this claims that it would be possible to factor a 2048-bit RSA number using 20 million noisy qubits in 8 hours. Their assumptions for this are: ``a planar grid of qubits with nearest-neighbor connectivity, a characteristic physical gate error rate of $10^{-3}$, a surface code cycle time of 1 microsecond, and
a reaction time of 10 microseconds''. See \cite[Appendix B]{GidneyEkera2021a} for details, but we note that these times are thousands of times slower than individual operations on a classical computer. Note that \cite{GouzienSangouard2021a} says 177 days, rather than 8 hours, which shows the uncertainty!

A different kind of investigation of Shor's algorithm is in \cite{Yamaguchietal2023a}, based on Fujitsu's mpiQulacs simulator (of perfect devices). This concludes that a 2048-bit RSA number would need 10241 qubits, with $2.22 \times 10^{12}$ gates, and depth $1.79 \times 10^{12}$.

\subsection{A pragmatic approach}\label{sec:Bernstein}

At the end of section \ref{sec:NFS}, we observed that there was a balance between the cost of searching and the cost of linear equation solving. But Grover's algorithm greatly reduces the cost of searching, essentially searching over $n$ objects for an object that satisfies a quantum oracle in time $O{\sqrt n}$, using essentially just the qubits required for the quantum oracle.

\cite{bernstein2017low} makes use of Grover's algorithm, suggesting that we use this to sieve much larger areas, and hence allowing a smaller $B$.
More precisely, in the GNFS algorithm, we divide the search area into tiles $(a,b) \in [a_0,a_1]\times[b_0,b_1]$ which are expected to contain \emph{precisely} one pair $(a,b)$ satisfying the smoothness criterion. 
In the terminology of \cite{schulz2022accelerating}, we are \emph{offloading} these tiles to a quantum engine. \cite{bernstein2017low} uses Shor's algorithm \emph{in quantum superposition} to check the smoothness criterion. Ongoing research at Bath is looking at alternatives here.

\cite{bernstein2017low} gives this formula for the time complexity of factoring $N$, an $n$-bit integer.
\begin{equation}\label{eq:Bernstein}
\exp\left(\left((8/3)^{1/3}+o(1)\right)(\log N)^{1/3}(\log\log N)^{2/3}\right).
\end{equation}
Note that $(8/3)^{1/3}\approx 1.387$. \cite{bernstein2017low} states that the number of qubits required is $O\left(n^{2/3}\right)$, which is clearly asymptotically better than Shor. \cite{Thorne2020a} does a more precise analysis, and shows
\begin{equation}\label{eq:BernsteinQ}
8\left(n^{2/3}\right)+4\left(n^{2/3}\right)\log(1.44\left(\left(n^{2/3}\right)\right).
\end{equation}
If a tile returns $\bottom$ (failure), it is not obvious whether
\begin{enumerate}
    \item[(a)]there is no suitable $(a,b)$ pair in the tile;
    \item[(b)]there are more than one suitable pairs, and Grover's algorithm has therefore failed to converge.
    \item[(c)]the quantum implementation failed (due to errors) and actually precisely one suitable $(a,b)$ pair exists.
\end{enumerate}
There is some treatment of this issue in \cite[\S22.1.4]{Aaronson2021b}, but this has not been integrated with the \cite{bernstein2017low} approach yet.
More practical experience with Grover's algorithm is clearly needed.

How likely is it that a tile will have precisely one suitable pair? If we assume (as we always do in GNFS studies) that the distribution of pairs can be treated as random, then Poisson distribution theory tells us that this probability is $1/e\approx 0.368$. This is ``only a constant factor'', but again more engineering and experience is required. Note that the probability of two results is $1/2e\approx 0.184$, and \cite[\S22.1.4]{Aaronson2021b} states that, in this case a run of Grover's algorithm for $1/\sqrt2$ as long should recover \emph{one of these}. Searching where the number of solutions is $>2$ is also possible \cite[Theorem 3]{Boyeretal1998}, but not necessarily efficient, as this event has probability $1-\frac1e-\frac1e-\frac1{2e}\approx 0.08$.

Note that, if a quantum search over a tile returns a result (rather than $\bottom$), it is trivial to check the results classically, and indeed we have to, as the method of \cite{bernstein2017low} merely returns the $(a,b)$ pair. It is worth noting that this is an intrinsic property of the inability to save state or break down quantum algorithms as one could with a classical algorithm.

\subsection{Comparison}

If $n$ is the number of bits in the number $N$ we are aiming to factor, let $L_{1/3}:= \exp\left((\log N)^{1/3}(\log\log N)^{2/3}\right)$.
Let $\gamma=(8/3)^{1/3}\approx 1.387$. 

\begin{table}[h]
    \centering
    \caption{Comparison of algorithms}
    \begin{tabular}{c|c|c}
Algorithm&Time &qubits \\
GNFS \S\ref{sec:NFS}  &(\ref{eq:GNFS})=$L_{1/3}^{\gamma^2+o(1)}$ & 0\\
Shor \S\ref{sec:Shor}  &$O(n^3)$ & $2n$\\
Hybrid \S\ref{sec:Bernstein}   &(\ref{eq:Bernstein})=$L_{1/3}^{\gamma+o(1)}$ & (\ref{eq:BernsteinQ})=$O\left(n^{2/3}\log n^{2/3}\right)$\\
    \end{tabular}
    \label{tab:my_label}
\end{table}
Then the above complexity theory can be summarised as Table \ref{tab:my_label}.
If (and this is dubious in practice) we ignore the $o(1)$ terms, we see a very substantial reduction in exponent of $L_{1/3}$ from GNFS to the hybrid method of \S\ref{sec:Bernstein}. But of course Shor is much faster.  Does \S\ref{sec:Bernstein} use fewer qubits than Shor? Asymptotically it clearly does, for small $n$ it clearly doesn't, so where is the transition point? Alas \cite{Thorne2020a} shows this as $n\approx4\cdot10^5$, whereas current values of $n$ are typically in the range 1024--8192. However, we should emphasise that \cite{Thorne2020a} is, as far as we know, the first attempt to quantify the $O\left(n^{2/3}\right)$ in \cite{bernstein2017low}, and further research is going on at Bath to improve this.
%

\section{The Missing Gaps}

Despite decades of research and staggering leaps forward, particularly in recent years, quantum computation is still in its infancy and it will require a great deal of work to bring quantum devices into widespread acceptance and use within the current scientific computing community and beyond.  This work must focus not only on the hardware advances, important though they are, but on what is required for an existing supercomputer centre to host, maintain and support the use of quantum devices as a part of their HPC service.  Critically, support of users, especially at the ``pump priming'' phase of early introduction to scientific computing, can only be effective if the training courses and software environment are well developed.  This will require a substantial investment, on top of any infrastructure changes and the cost of the system itself.

\subsection{The people} 

Many graduates in the late 1980's, 1990's would leave University with an understanding of programming, being able to write applications in FORTRAN, COBOL, Pascal or even C.
It may not have been optimal code but with further experience at work and training in optimisation and programming techniques some highly optimised codes were developed.  
When massively parallel systems became the norm, further development was needed to learn MPI\cite{mpi40standard} or UPC\cite{upc13standard} \textemdash undergraduate courses were also moving to include some of these new features
in lecture notes.  Now that large multi-core systems are ubiquitous, many STEM courses have moved to introduce students to parallel programming and domain decomposition, and most courses with some programming content will cover the basics of techniques such as multi-threading.

In stark contrast to this, there is a limited number of graduates who leave university with any experience of programming quantum computers.  It is a completely different programming paradigm, in 
some limited respect analogues to moving from serial to parallel applications, but much more complex.  \cite{quantuminsider} provides a list of the top 5 
quantum programming languages and gives a good explanation of the difference in understanding required to write quantum software.

The result is that, while it is not too difficult to recruit programmers and software developers who understand how to exploit parallelism to at least some degree, experienced quantum programmers are rare indeed.  Quantum circuit simulators, such as Qiskit \cite{qiskit} and qsim \cite{qsim}, can help interested programmers to build experience without access to quantum hardware.  However, the steep learning curve, perhaps better termed a ``learning cliff'', presents a significant barrier to many.  There is also a lack of knowledge in the general HPC community of how best to approach such a fundamentally different hardware architecture, and how quantum computation might benefit scientific applications.

\subsection{The Software}
Debuggers, IDEs, optimising compilers and profilers are now essential tools for the modern day software developer, software maintainer and programmer.  In quantum computing however, most of these currently have no analogue, and those that do are far less mature than their classical cousins.  Programming of quantum devices requires a radically different approach, and consequently the tools required are very different to those the HPC community has so far been accustomed to.  \cite{schulz2022accelerating} attempted to lay out a vision for how the HPC community might move as smoothly as possible to exploiting quantum devices, but admit that there is some considerable work ahead to make this a reality.

There are also other, previously mentioned challenges that must be addressed, specifically quantum error detection and handling, be that by mitigation or correction, which are likely to be themselves error prone when implemented by less experienced quantum programmers.  They may also prevent other programmers from attempting the move to quantum hardware.  This is certainly an area where improved tooling, such as \cite{ball2021Software}, and software middle-layers such as \cite{riverlaneDeltaflowOS}, can help.  Several of these make use of logical, rather than physical, qubits to shield the user and programmer from the pain of having to mitigate the huge amount of noise and errors that they would otherwise encounter from the underlying hardware.  Logical qubits can be thought of as a type of quantum simulation of high-fidelity qubits run on noisy, error-prone quantum hardware.  This sounds like an obvious solution, but logical qubits can only exist once we have a fault-tolerant error correction architecture.  See \cite{gambetta2017building} for a more detailed description of logical qubits, and how one might use them to create a quantum memory.
%

\section{Energy, Maintenance and Infrastructure}
It is well documented that the worlds fastest supercomputer (Frontier) consumes 21MW of power topping the current TOP500 \cite{top500} with an impressive 1.102 exaFLOPS.  There is, currently, no direct comparison
available for quantum computers at scale.  
Again it is well publicised that, for today's quantum computers, the majority of power (20kW -- 30kW) goes to keep the system at optimal operating 
temperature, perhaps 15 millikelvin (-273\degree{C}), and when applications run they add, on average, 1 - 2 kW to the power budget.  Due to the current super-cooled designs of many quantum computers, the
cooling capacity should need little modification to expand and allow for larger qubit counts.  However, there is no information available that shows how
power consumption will increase when the number of qubits goes beyond what the current infrastructure allows.

A wide variety of quantum hardware implementations are currently being investigated, several of which are described in \cite{enriquez2023estimating}, and until this converges we cannot begin to speculate on the difficulties of planning maintenance sessions.  Most are using super-cooled chambers, which can take several days to get to the correct operating temperature following maintenance, on top of the time taken to warm up to a safe temperature for the maintenance itself to take place.  This cooling requires a considerable amount of energy, and is beyond the infrastructure currently installed in the majority of data centres.

With modern conventional processors it is possible to limit the
processor speed, or not buy the fastest chips\footnote{For example, JHD did this for the University of Bath's purchase: 2.8GHz chips were less that 10\% faster than 2.6GHz, but consumed 25\% more power.}, therefore reducing the power consumption of the system.  There are early indicators that large HPC installations are re-evaluating their ``green credentials'' and noting that significant savings on energy costs can be made with throttling back processors by a few percent\cite{archer2} - clearly this will affect applications in different ways. 
CPUs also save power by powering off parts of the chip that are not currently experiencing demand, such as the vector registers.
At the current time, it is not clear how such a mechanism could work with quantum devices, or whether it might be possible to reduce power usage by, for example, powering down all or part of a device when it is not in demand.

Arguments suggesting that quantum computation is ``greener'' than classical computation tend to be based solely on the energy efficiency cost of the algorithms themselves.  This gives no indication of the increase in carbon emissions due to the cooling infrastructure, as the hardware designs seemingly most likely to be deployed in a production setting are super-cooled.  Nor do they take into account the warming/cooling cycle during what are likely to be frequent maintenance periods, nor do calculations for either consider the emissions caused by the shipping of parts. This is not to say that quantum \emph{might} not be greener: merely that we can't do these calculations until we actually have the production machines.

In large part this is because no one yet knows quite how reliable production quantum devices will be, making it hard, if not impossible, to guess at the likely maintenance requirements.  At this time it is hard to state with any certainty that a production-ready quantum computer will be greener than current state of the art HPC systems.  What is more likely is that, once the teething problems are overcome and maintenance requirements are understood, a \emph{mature} quantum-enabled supercomputer will deliver significantly more FLOPS per Watt than its current non-quantum counterpart.  This however depends on both users and software adapting to take advantage of the new devices, and that again will take time.  Early systems are unlikely to be very green if their entire hardware ecosystem and reliability are taken into account, especially if they are often left idle.

The design of modern HPC systems also allows for degraded running.  If a processor or memory DIMM fails it is possible to repair this ``on the fly'' without the need 
to take the entire system offline.  Currently the design of the quantum computer does not lend itself to this controlled repair cycle.  Any repairs will require the 
system to be returned to room temperature, the problem addressed, and finally cooled back to -273\degree{C}.  This warming and cooling process can take several days to complete,
time which cannot be used for computation.  As the modern day HPC uses commodity parts, and a relatively new technician can repair many of the common problems; the same is 
not true for quantum systems, partly due to the cooling infrastructure.

\section{Conclusion}
We have seen that quantum computing, as well as being different in the small (as, say, GPU programming is different from CPU computing), is also different in the large. Not only do we not have large, error-prone for that is the nature of computing with quanta, computers, we don't really know what to do with them if we had them, how we would program them, or who, at scale, could do this.

We have also seen that the hardware demands of current quantum devices are, as with all new technologies, greater than that of current classical hardware.  Maintenance of a device normally running at -273\degree C is accordingly more challenging and will need careful planning, as well as a great deal of additional training of support staff and field engineers.  All early technologies have a tendency towards a greater frequency of hardware faults, leading to more frequent maintenance intervals to ensure that any production system meets its minimum operational capacity.  Maintenance plans, as well as the sort of parts logistics (and preemptive part ordering prediction) which are already needed for classical exascale computers, will have to be adapted to meet these new challenges.  At the present time it seems likely that any early quantum computer will require considerable operational expenditure.

It seems unlikely that \emph{early} quantum computers will be able to deliver on the green promises that are being made.  Nevertheless, in the longer term large savings are expected due to the significant increase in computational throughput.

\section*{Acknowledgements}
The authors thank Matthew Thorne for \cite{Thorne2020a} and many discussions with JHD, and Ben Pring for useful comments on a draft.
\bibliographystyle{unsrt}
\bibliography{main}
\end{document}